\documentclass[12pt]{article}
\usepackage{amssymb,amsmath,mathrsfs}
\usepackage{hyperref}
\usepackage{graphicx}
\usepackage{color}
\usepackage[OT2,OT1]{fontenc}
\usepackage{upquote}
\usepackage{float}
\usepackage[margin=1cm]{caption}
\usepackage{authblk}
\usepackage[numbers,sort&compress]{natbib}

\usepackage[labelfont=bf,labelsep=period]{caption}

\begin{document}

\title{\vspace{-1cm}\Large\bf Efficient application of the Radiance Enhancement method for detection of the forest fires due to combustion- \\
originated reflectance}

\author[1, 2, 3]{\small Rehan Siddiqui}
\author[1, 3]{\small Rajinder K. Jagpal}
\author[1, 2, 4]{\small Sanjar M. Abrarov}
\author[2, 3, 4]{\small \\ Brendan M. Quine}

\affil[1]{\scriptsize Epic College of Technology, 5670 McAdam Rd., Mississauga, Canada, L4Z 1T2 \normalsize}
\affil[2]{\scriptsize Dept. Earth and Space Science and Engineering, York University, 4700 Keele St., Canada, M3J 1P3 \normalsize}
\affil[3]{\scriptsize Dept. Physics and Astronomy, York University, 4700 Keele St., Toronto, Canada, M3J 1P3 \normalsize}
\affil[4]{\scriptsize Thoth Technology Inc., Algonquin Radio Observatory, Achray Road, RR6, Pembroke, Ontario, Canada, K8A 6W7 \normalsize}

\date{October 6, 2022}
\maketitle

\vspace{-0.75cm}

\begin{abstract}
The existing methods for detection of the cloud scenes are applied at relatively small spectral range within shortwave upwelling radiative wavelength flux. We have reported a new method for detection of the cloud scenes based on the Radiance Enhancement (RE). This method can be used to cover a significantly wider spectral range from 1100 nm to 1700 nm by using datasets from the space-orbiting micro‑spectrometer Argus 1000. Due to high sunlight reflection of the smoke originated from the forest or field fires the proposed RE method can also be implemented for detection of combustion aerosols. This approach can be a promising technique for efficient detection and continuous monitor of the seasonal forest and field fires. To the best of our knowledge this is the first report showing how a cloud method can be generalized for efficient detection of the forest fires due to combustion-originated reflectance.
\vspace{0.25cm}
\end{abstract}

\section{Introduction}

Increase of carbon dioxide gas appearing as a result of intense production of goods in industrial and agricultural sectors of economy is a main issue of the modern human society that causes uncontrollable rise of atmospheric temperature due to devastating greenhouse effect. In particular, just in five decades the concentration of $\rm{CO_2}$ has been rapidly raised from 288 ppm to more than 410 ppm in 2020 \cite{Apadula2019, Karnauskas2020, Ueyama2020}. As a consequence, the global warming of the atmosphere variates the weather dynamics causing tremendous negative impact to the flora and fauna of the Earth \cite{Scarpa2020}.

The importance of clouds and their significant roles in sustaining the temperature balance on the Earth cannot be overestimated \cite{Guo2015, Fournier2006, Mitchell2009, Siddiqui2015}. However, a stable positive rate 2.05 ±0.03 ppm/year of concentration of carbon dioxide greenhouse gas over the last decades increases the atmospheric temperature causing more intense water evaporation and formation of clouds \cite{Apadula2019}. Consequently, we observe more thunderstorms that drastically increase the chances for ignition of the forest fires due to lightning. As a result of global warming, the annual forest fire season in Canada has been extended from April to November \cite{Tymstra2020}. Furthermore, a rapidly increasing population of Canada greatly intensifies tourism and hunting that ultimately causes forest fire ignitions due to absolutely uncontrollable conditions of camping. There were also reports that some arsonists can ignite forest fires purposely \cite{Ganteaume2013, Axelson2009}.

Forest fires mostly burn down and destroy the pine trees in the North America. Pine trees grow extremely slow and, therefore, unlike many other species their recovery process may take many decades. The cold weather and clean environment are vital factors for their growth. However, increase of atmospheric temperature and anthropogenic pollutants significantly reduces chances for their survival. Consequently, the pine trees are gradually replaced by other species mainly by broadleaf trees and bushes. Pine trees are flammable and very vulnerable to heat. They are necessary sources of high quality wood that are used to build houses, furniture and electric poles to carry electricity. There are very limited resources that could replace pine trees and none of them are more cost-effective and better in quality. Therefore, losing these valuable pine forests may negatively affect the economy in the future \cite{Parisien2020}.

The seasonal forest fires can last for many months and produce a large quantity of the smoke that endangers many animals by destroying their natural habitats \cite{Parisien2020, Tymstra2020, Holm2021}. Furthermore, according to the recent studies as a result of large areal coverage and prolonged exposure of the smokes containing hazardous aerosols, the forest fires can cause the various respiratory deceases including the lung cancer among local farmers in rural areas situated close to forests \cite{Wind2010}. Therefore, detection and timely prevention of the forest fire spreads are highly needed.

Alongside with conventional methods based on MODIS and Google cloud database that provides precise and efficient probabilistic approach in determination of the multi-layered cloud masking \cite{Wind2010, Tang2013}, the infra-red (IR) remote sensing is another approach for detection of clouds \cite{Siddiqui2015, Siddiqui2017a, Siddiqui2017b, Siddiqui2020a}. In particular, in our recent publications we have described a new method for detection of cloud scenes based on the Radiance Enhancement (RE) method \cite{Siddiqui2017b, Siddiqui2020a}. Despite some drawbacks of the RE method such as lower resolution and difficulties to distinguish multi-layered clouds, its application, nevertheless, may be advantageous especially when remote sensing is provided from the space-orbiting IR micro-spectrometer like Argus 1000 \cite{Siddiqui2017b, Siddiqui2020a}.

\begin{figure}[H]
\centering
\includegraphics[width=20pc]{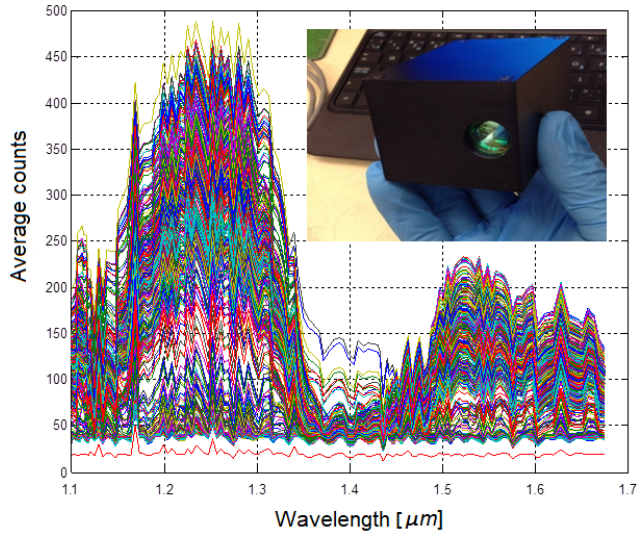}
\caption{\sffamily The space observation package of Argus 1000. Inset shows a photograph of the light and small-size Argus 1000 micro-spectrometer.}\label{fig1}
\end{figure}

The ultra-light (about 0.25 kg only), small-size and inexpensive Argus 1000 instrument was launched into space in 2008 from India as a payload of the CanX-2 nanosatellite \cite{Rankin2005, Jagpal2011}. Being in space it provides the large-scale datasets as shown in Fig. \ref{fig1} that can be used to extract valuable information about atmospheric gas constituents like concentration of $\rm{CO_2}$ and other greenhouse gases by retrieving of the IR spectral radiance \cite{Jagpal2011, Jagpal2010, Jagpal2019}. Inset in Fig. \ref{fig1} demonstrates a photo of the Argus 1000 micro-spectrometer.

Apart from determination of $\rm{CO_2}$ and other greenhouse gases, the IR radiance data collected by Argus instrument from space enable us to develop the RE method for efficient detection of the cloud scenes. Recently we suggested that a detection method of the cloud scenes can be generalized for detection of the forest and field fires \cite{Siddiqui2020a, Siddiqui2020b}.

\begin{figure}[H]
\centering
\includegraphics[width=20pc]{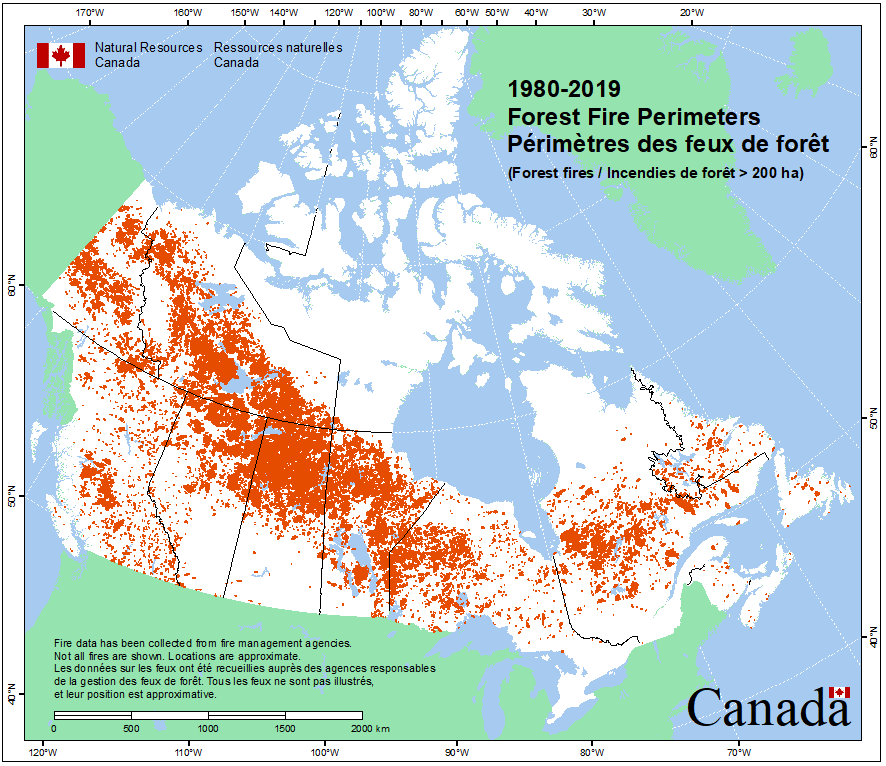}\\
\caption{\sffamily Map of Canada with forest fire locations during the period from 1980 to 2019 \cite{NRC}.}\label{fig2}
\end{figure}

Figure \ref{fig2} shows a map of Canada with forest fire locations depicted by the red spots \cite{NRC}. In average a typical fire in a forest covers the range of a few squared kilometers only. However, its impact to the forest is much devastating. In particular, the forest fire causes the smoke that extends for tens and hundreds of squared kilometers increasing the temperature, spoiling air quality and preventing the sunlight locally for several months. Taking into consideration coverage areas of the spreading smokes, one can estimate from this map that about two thirds of the forest territories in Canada have been affected by forest fires during 1980 to 2019 years \cite{NRC}. The deficiency of the sunlight, pollution occurring due to smokes and an increased temperature of the environment can kill pine trees even in a larger quantity than the actual fire that burns them down. As a result, being unaffected directly by a fire these trees retain all material qualities including high density of wood and beam length. Consequently, these dead pine trees with high quality of wood are logged in a commercially large scale.

Conventional methods for forest fire detection include video-cameras, thermal imaging cameras, IR spectrometers to identify the spectral characteristics of smoke, light detection and ranging systems (LIDAR) and so on \cite{Alkhatib2014, Wang2013, Benali2016}. However, in contrast to the conventional methods, the RE technique has some advantages such as global coverage and continuous monitor due to periodic observation from the space-orbiting nanosatellite in the real time mode. Another important feature is that the RE method can cover a wide spectral wavelength range to detect enhancement of flux from the surface especially due to clouds \cite{Siddiqui2017b, Siddiqui2020a}.

In this work we show how the RE technique can be used to detect efficiently the forest fires by using Argus 1000 space data. The main objectives of this study are to implement the RE to the line-by-line (LBL) forward radiative transfer model that accounts for wavelength dependency of surface reflectance with method of weighted sum and to show numerically a consistency between space observation and computed radiance spectra. Our method is based on a match between synthetic and observed radiance that accounts for high surface reflectance appearing as a result of smokes produced by forest fires. A generalization of any water vapor cloud methods for detection of the forest fires has never been reported in scientific literature.

\section{Methodology}

The Radiance Enhancement methodology is based on least square match between space observation and synthetic datasets. In order to generate synthetic data we used LBL radiative transfer model GENSPECT \cite{Quine2002}. This forward model computes the radiance for the nadir and limb observations for the greenhouse gases with all required parameters provided by the HITRAN molecular spectroscopic database \cite{Hill2016}. Some supplementary MATLAB files are additionally developed to improve performance of the model. In particular, for computation of the absorption coefficients we applied a newly modified code for more rapid and accurate computation of the Voigt function based on a new single-domain interpolation technique \cite{Abrarov2019} for which the highly accurate reference values can be generated by using any of three rapid algorithms described in our works \cite{Abrarov2011}, \cite{Abrarov2018a} or \cite{Abrarov2018b}. In contrast to the conventional algorithms \cite{Fomin1995, Sparks1997}, in our implementation we interpolate the Voigt function in a single domain itself in order to avoid unnecessary interpolation in computation of the absorption coefficients.

The model GENSPECT accounts for different variables that include concentrations of the greenhouse gases, deviation of the nadir angle, zenith angle of sun, and so on \cite{Quine2002}. In the latest updates we developed some function files that also account for wavelength dependency of the reflectance due to different surface Albedo like clouds, pine-trees, vegetation and grass \cite{Roberts2013, Li2018, MODIS, Baldridge2009}. The radiative transfer model is run in a nested loop by incrementing/decrementing values of the fitting variables until a best match is achieved.

The RE methodology is based on the following formula \cite{Siddiqui2017b, Siddiqui2020a}
\[
R{E_i} = \frac{1}{N}\sum\limits_{j = 1}^J {\left\{ {\frac{{OBS_i\left[ j \right] - SYN_i\left[ j \right]}}{{SYN_i\left[ j \right]}}} \right\}},
\]
where $i$ is the index of wavelength sub-bands, $j$ is the index of grid-points, $J$ is the total number of the grid-points and $N$ is the number of sub-bands that can be taken as $4$. This methodology shows the efficiency in determination of the cloud scenes. The corresponding Combined Radiance Enhancement (CRE) formula is given by \cite{Siddiqui2017b, Siddiqui2020a}
\[
CRE = \sum\limits_{i = 1}^N {R{E_i}}.
\]

The RE and CRE values can be used to predict the cloud scenes. If the surface Albedo is relatively high, say above 0.6, then the best match between observed and synthetic radiance signifies that the specific observation is likely due to thick cloud or any scattering particles such as ice pallets or aerosol. In this case we should expect that the RE and CRE values to be small by absolute values. Specifically, when CRE is relatively close to zero at high reflectance, we can expect higher chances for cloud scene for a specific location due to high surface Albedo. The CRE is predefined for the wavelength bands and accounts for concentration of the selected greenhouse gases \cite{Siddiqui2017b, Siddiqui2020a}.

Although the RE method cannot distinct the cloud scenes with aerosols in from of solid particles and liquid droplets (including particulate matters $\rm{PM_{2.5}}$ \cite{Christopher2006}), it, nevertheless, can be advantageous in practical applications. If the weather forecast predicts no presence of clouds while the RE method shows their availability, then we can conclude that these type of reflectance could be due to aerosols like dust \cite{Mamun2021}, industrial pollutants from big plants, hydroelectric stations or, more likely, combustion due to seasonal forest fires that typically produce a large amount of smokes. Consequently, as a result of high reflectance of the combustion-based aerosols, the RE method can also be used efficiently for detection of the forest fires.

\begin{figure}[H]
\centering
\includegraphics[width=20pc]{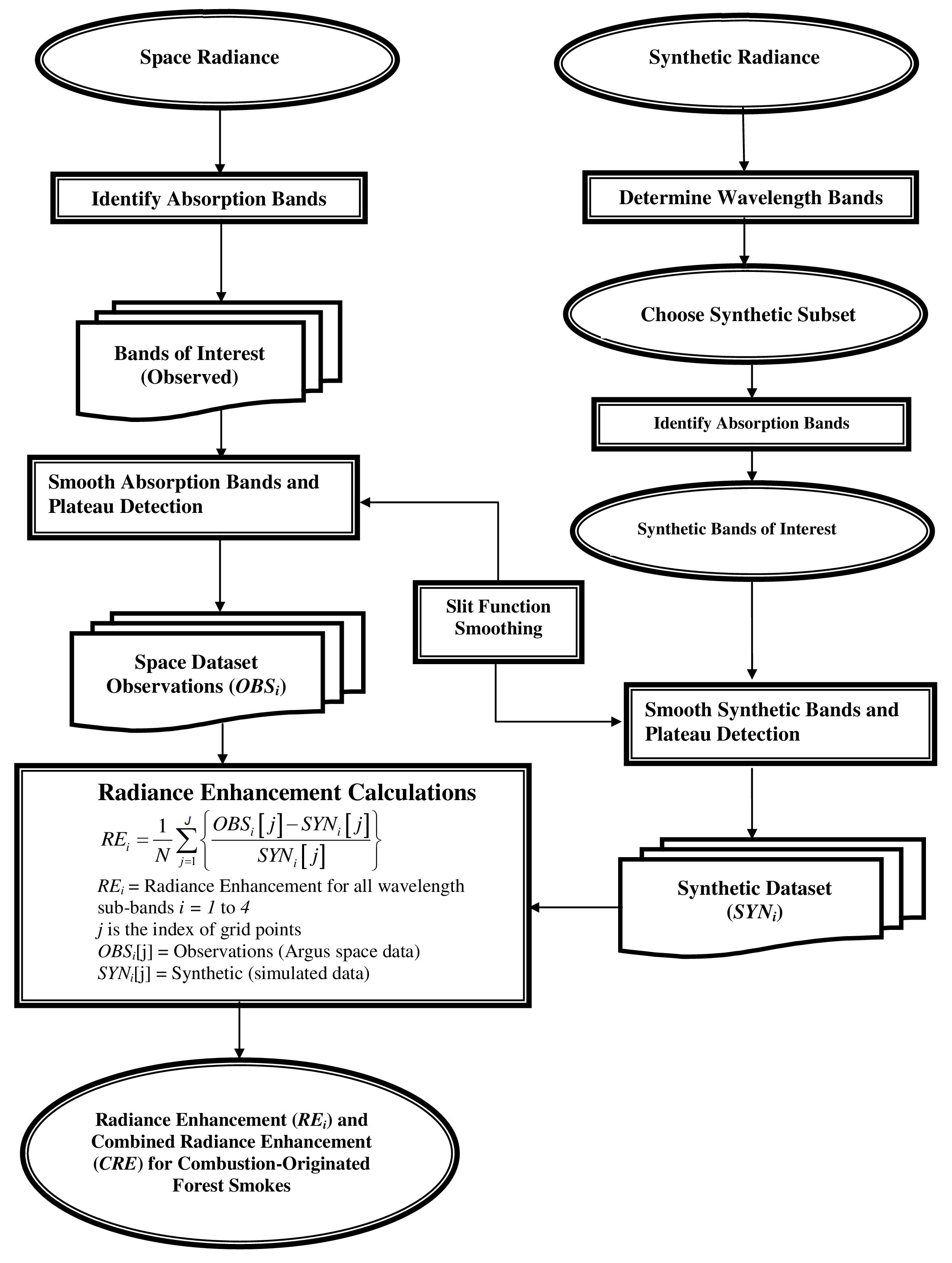}
\caption{\sffamily Flow-chart for computation of the RE method.}\label{fig3}
\end{figure}

Figure \ref{fig3} shows flow-chart of the RE method for detection of the aerosol cloud due to forest fire that compares space observation and synthetic data. As we can see from this figure, in the intermediate stages the synthetic data is also divided into four sub-bands and passed to the slit function smoothing that simulates resolution of the Argus instrument. Once the RE computation is completed, the comparison of the space observation and synthetic data is performed and if there is a best match, then the corresponding observation data is chosen. The detailed description of the RE algorithmic implementation can be found in the work \cite{Siddiqui2017b}.

\section{Results and discussion}

\begin{figure}[H]
\centering
\includegraphics[width=24pc]{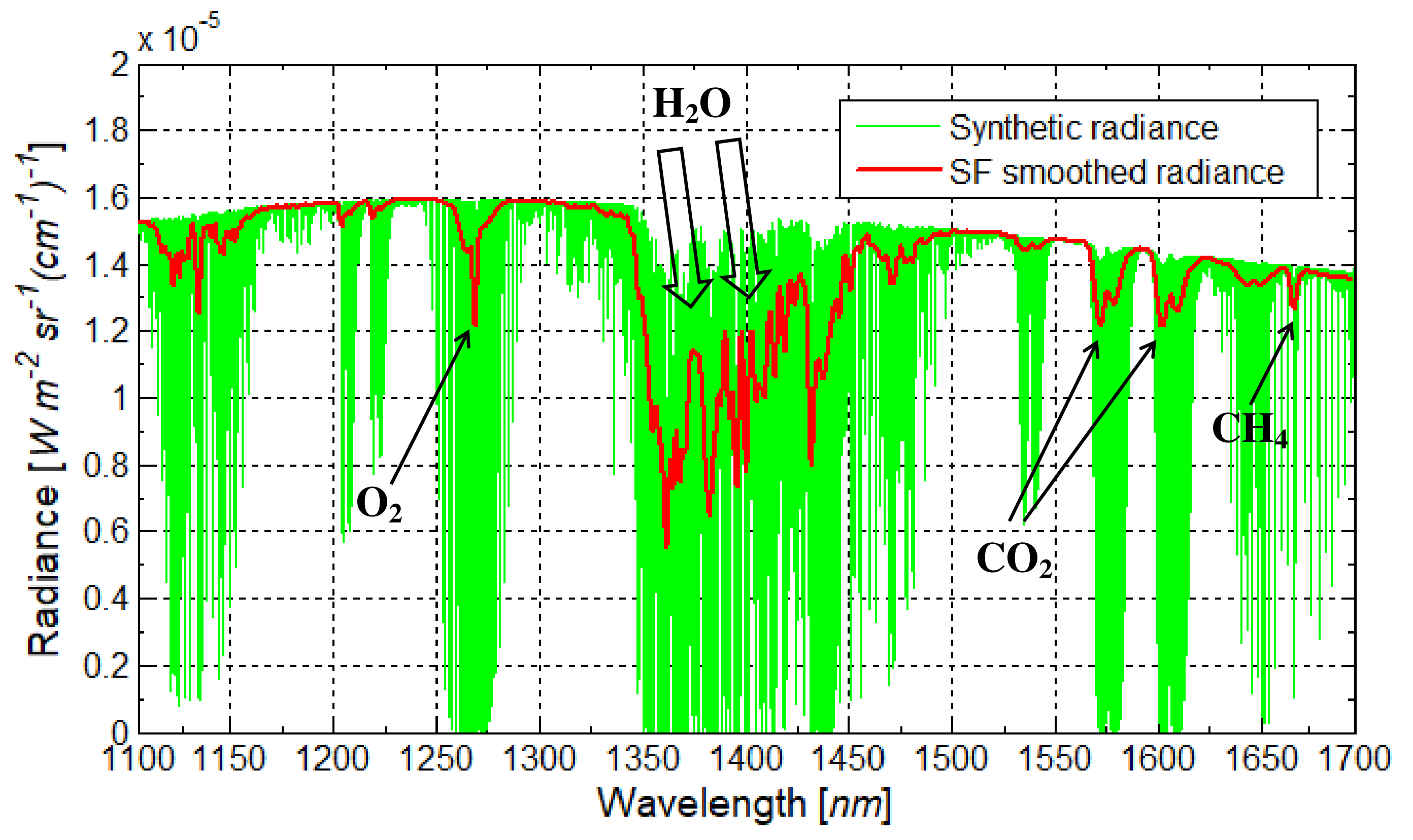}
\caption{\sffamily Synthetic radiance computed by radiative transfer model GENSPECT \cite{Quine2002} at a constant reflectance.}\label{fig4}
\end{figure}

The LBL radiative transfer tool GENSPECT \cite{Quine2002} generates synthetic spectral radiance for the greenhouse gases that can be used for comparison with space observation data. Figure \ref{fig4} shows the synthetic spectral radiance for $\rm{CO_2, H_2O, O_2}$ and $\rm{CH_4}$ gases computed at constant reflectance. The green shadow area is originally computed radiance while the red curve depicts the slit function smoothed radiance that simulates the limited resolution of the space instrument. The arrows in this figure indicate absorption band due to $\rm{O_2}$ near 1260 nm, the wider band of absorptions due to water vapors from 1300 nm to 1480 nm, two absorption bands near 1575 nm and 1600 nm are due to $\rm{CO_2}$ and one narrow absorption band near 1650 nm due to $\rm{CH_4}$ \cite{Jagpal2011}. Each of these wavelength bands can be used in the RE method for the comparison with Argus observation data in order to detect cloud scenes and forest fires. Concentrations of $\rm{H_2O}$ and $\rm{CO_2}$ bands are especially essential variables for matching of the synthetic model with space dataset from Argus 1000 micro-spectrometer.

\begin{figure}[H]
\centering
\includegraphics[width=20pc]{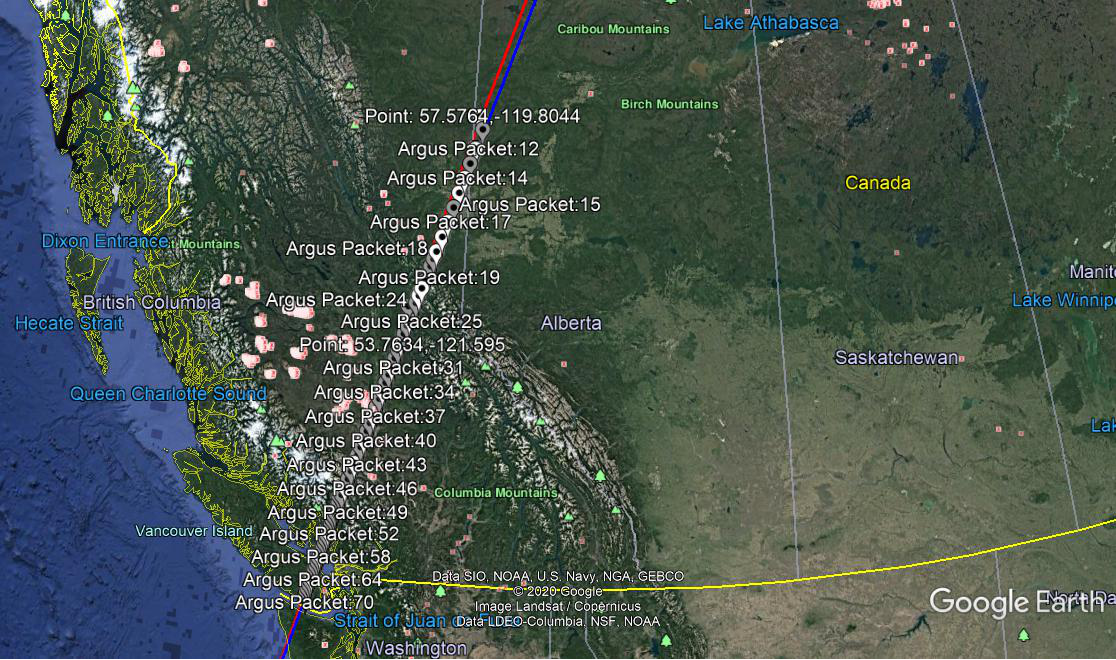}
\caption{\sffamily Geolocation of the Argus 1000 micro-spectrometer corresponding to week 11, pass 69 and observation numbers from 12 to 70, over British Columbia, Canada.}\label{fig5}
\end{figure}

The geolocation of the Argus instrument has been determined with help of Systems Tool Kit (STK) \cite{STK}, \cite{GE} and \cite{EOSDIS} for location of the forest fires over Canada. The datasets from 2009 to 2015 for Argus instrument has been processed and analysed for water vapour and combustion-based aerosol clouds. Figure \ref{fig5} shows Argus 1000 path, observation numbers from 12 to 70 for week 11 and pass 69, over British Columbia, Canada. As we can see from Fig. 5, the red spots indicating the actual forest fire locations intercept the trace of space orbiting IR remote sensor corresponding to observation packet numbers 12 to 52.

Increase of atmospheric temperature gradually changes the canopy constituents of the forests; the burned and dead pine trees are not necessarily replaced by new generation of pine trees. Due to increasing atmospheric temperature, the chances for spreading of broadleaf trees and bushes are higher since for successful competition of young generation of the pine trees with other species a colder environment is highly preferable. Nowadays we can observe explicitly that after seasonal forest fires in the Algonquin National Park, Ontario, Canada, the pine trees are replaced by broadleaf trees, bushes and grasses changing the appearance and nature of the forest. Moreover, with development of the agricultural sector the more and larger territories of the forests are transformed to prairies, which are intensively used now by farmers for grazing cows, lambs, goats and horses. The density of trees in the forests is also declined due to increasing demand of most valuable wood of pine trees in the market and environmental pollutions.

\begin{figure}[H]
\centering
\includegraphics[width=24pc]{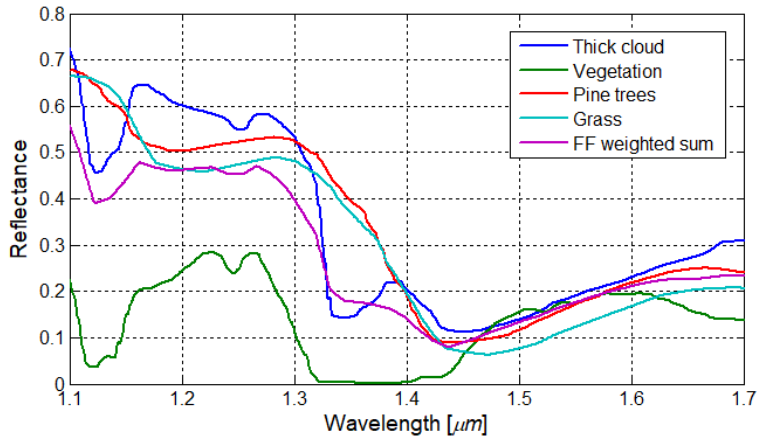}
\caption{\sffamily Reflectance dependencies for thick cloud, vegetation, pine trees, grass and their cumulative weighted sum.}\label{fig6}
\end{figure}

\begin{figure}[H]
\centering
\includegraphics[width=24pc]{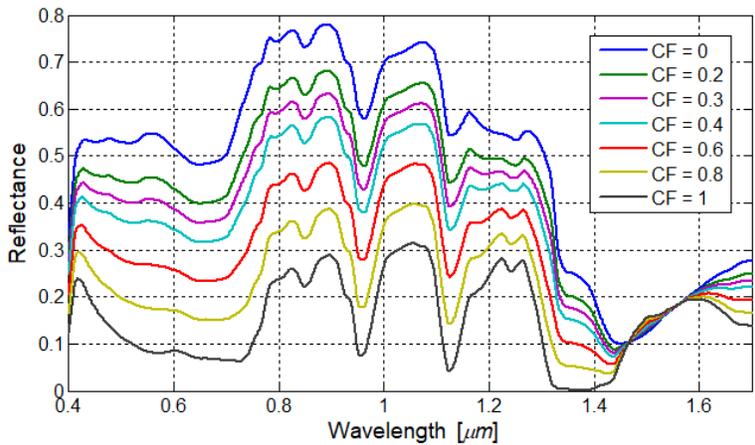}
\caption{\sffamily Evolution of the cumulative reflectance computed by weighted sum at different contribution factors.}\label{fig7}
\end{figure}

Surface reflectance of forests is generally the wavelength dependent rather than a constant. Therefore, a more rigorous consideration requires that the reflectance has to include cumulative contribution of cloud, pine trees, vegetation (broadleaf trees and bushes) and grass. The visual analysis of the forest in British Columbia suggests that pine trees occupy 0.5 to 0.7 of the ground area, while remaining area is occupied by broadleaf trees, bushes and grasses. Figure \ref{fig6} shows reflectance as a function of the wavelength for thick cloud, vegetation, pine trees and grass, which data can be obtained from \cite{Roberts2013, Li2018, MODIS, Baldridge2009}.

The forests can occasionally share the land with turbid rivers and lakes. However, it is relatively rare when forest fires occur in the neighborhood of water. Therefore, we do not consider these events in our model. As an example, Fig. \ref{fig7} illustrates the evolution of the reflectance occurring due to contribution factor (CF) from the vegetation. In our model we consider cumulative reflectance from the surface due to thick cloud, pine trees, vegetation and grass. While the contribution factor from the vegetation changes from 0 to 1 (as shown in Fig. \ref{fig7}), we imply that the relative ratios of the areal coverage corresponding to clouds, pine trees and grass remain same. Thus, the computation of cumulative reflectance was performed by weighted sum for all four types of surface reflectance.

\begin{figure}[H]
\centering
\includegraphics[width=24pc]{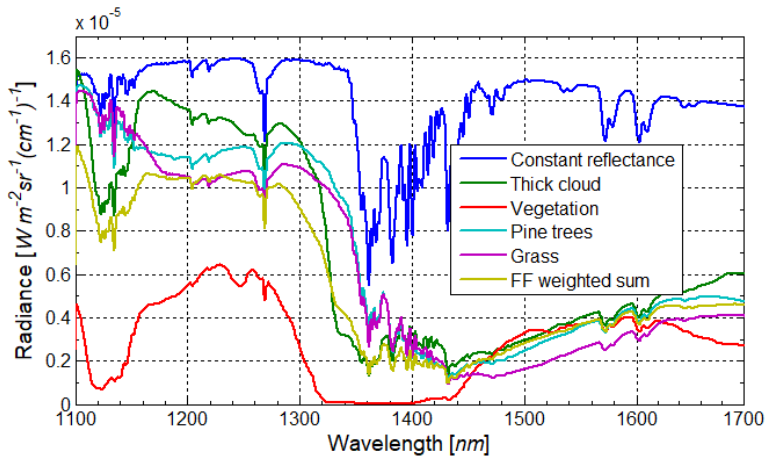}
\caption{\sffamily Synthetic radiance (all slit function smoothed) computed at different wavelength dependent reflectance.}\label{fig8}
\end{figure}

Figure \ref{fig8} shows the radiance computed at constant reflectance, thick cloud, vegetation, pine trees, grasses and by weighted sum due to forest fire (FF). As we can see from this figure, the wavelength dependency of the reflectance significantly changes the radiance. Particularly, the radiative transfer model GENSPECT \cite{Quine2002} generates the graphs where the right portion above 1400 nm is suppressed. This suppression effect on the right part of the spectral region plays an important role in retrieval process of the space observable data. Despite this one can still observe the profound absorption bands near 1575 nm and 1600 nm due to carbon dioxide greenhouse gas.

\begin{figure}[H]
\centering
\includegraphics[width=24pc]{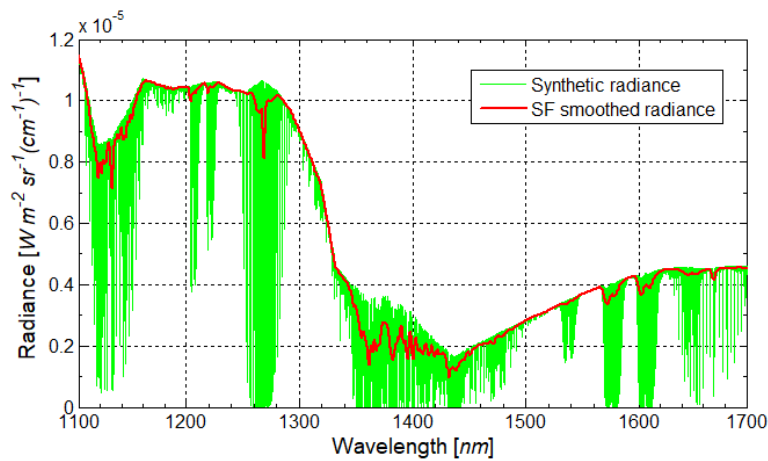}
\caption{\sffamily Synthetic radiance (smoothed) computed for cumulative wavelength dependent reflectance at CF = 0.3.}\label{fig9}
\end{figure}

\begin{figure}[H]
\centering
\includegraphics[width=24pc]{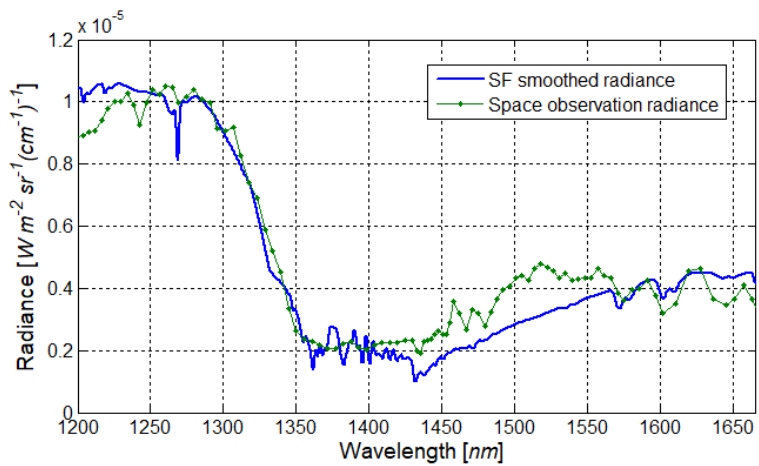}
\caption{\sffamily Comparison of synthetic radiance (smoothed) radiance and space observation radiance.}\label{fig10}
\end{figure}
 
Figure \ref{fig8} shows the synthetic radiance (green shadow area) and slit function smoothed radiance (red curve) that accounts for the wavelength dependent radiance. Comparing Figs. \ref{fig4} and \ref{fig9} with each other we can observe significant changes in the radiance. However, it should be noted that despite the suppression on the right portion of the graphs above 1400 nm, the two absorption bands of $\rm{CO_2}$ greenhouse gas near 1575 nm and 1600 nm still remain profound.

It has been found by experimental fitting that ${\rm{CF}} = 0.3$ provides the best match between synthetic and space observable data. This can be seen from the Fig. \ref{fig10} showing the space observation and synthetic radiance spectra. It should be noted that at 1575 nm and 1600 nm we can observe a relatively good match for carbon dioxide. The aerosol cloud due to forest fire and canopies makes a significant contribution in suppression of the radiance above 1400 nm. The concentration of $\rm{CO_2}$ in calculation is increased by 40\%. Alongside with the RE model, the increased level of $\rm{CO_2}$ also supports an assumption for presence of the forest fire. The synthetic spectrum shown in Fig. \ref{fig10} has been computed by incorporating all four different types of surface reflectance, specifically due to aerosol cloud, pine trees, vegetation (broadleaf trees and bushes) and grass. Space observation radiance spectrum in Fig. \ref{fig10} corresponds to Argus week 11, pass 69, observation number 49.

\begin{figure}[H]
\centering
\includegraphics[width=30pc]{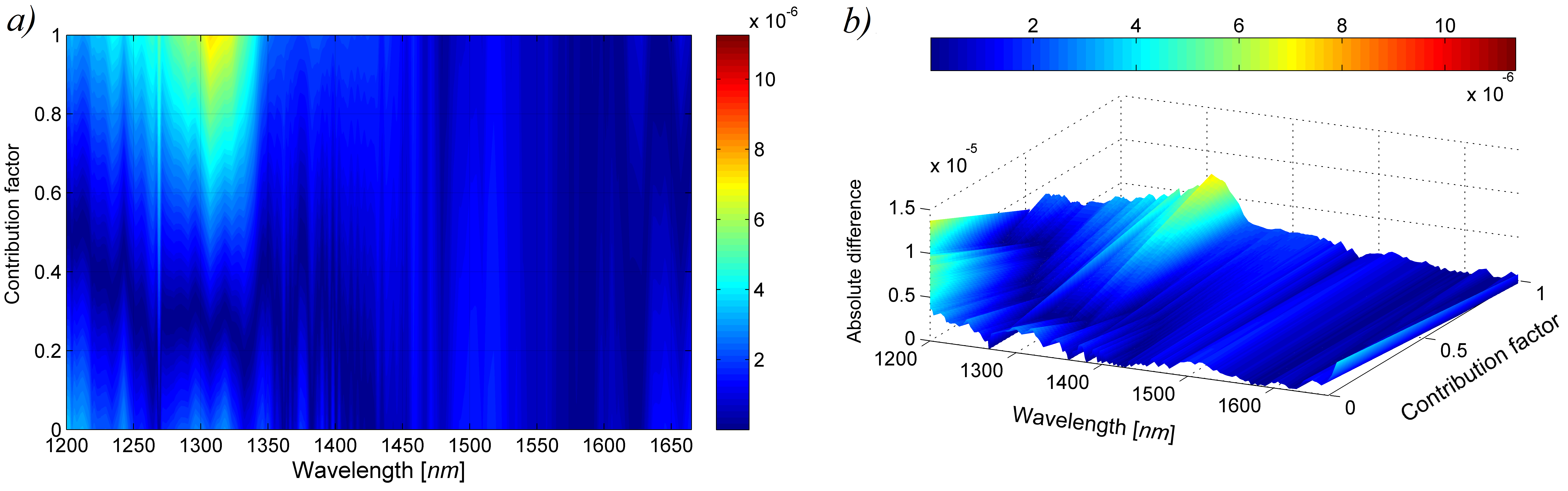}
\caption{\sffamily Absolute difference between synthetic and space observation data: a) 2D plot and b) 3D plot.}\label{fig11}
\end{figure}

Figures 11a and 11b show the absolute difference between synthetic and space observation data in 2D and 3D plots, respectively. As we can see from these figures, the darkest blue areas correspond to CF = 0.3. Change of vegetation CF higher than 0.6 or lower than 0.2 significantly increases the absolute difference. The weightages for the cloud, pine trees, vegetation (broadleaf plants and bushes) and grass are found to be 0.4, 0.3, 0.2 and 0.1, respectively. This signifies that corresponding areal coverage observed from space due to cloud, pine trees, vegetation and grass are about 40\%, 30\%, 20\% and 10\%. It should also be mentioned that as an alternative to the areal coverage, the weighted sum can also be performed by computing corresponding proportions from the upwelling radiative flux \cite{Siddiqui2017a}.

The error analysis for this RE method shows a reasonable agreement between synthetic and Argus space observation radiance to detect the forest fire location with given IR wavelength range.

\begin{figure}[H]
\centering
\includegraphics[width=24pc]{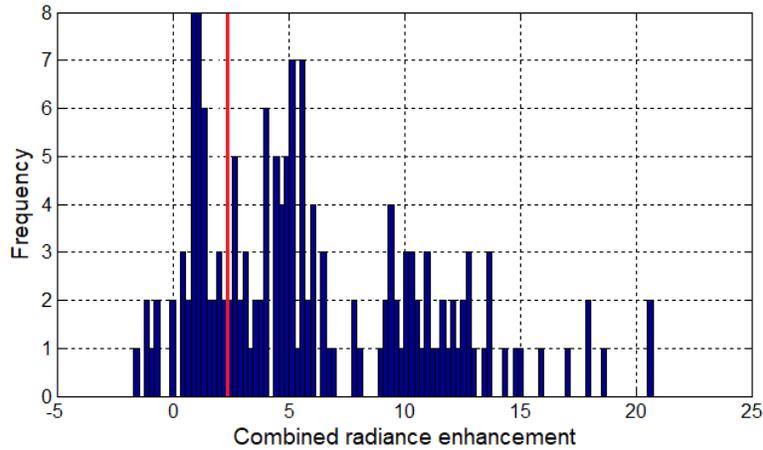}
\caption{\sffamily Frequency vs. CRE bar chart. The red line indicates the threshold value.}\label{fig12}
\end{figure}

The space observed flux generally gives enhanced radiance due to larger oblique view angles as compared with nadir view angles because of high clouds and aerosols clouds thickness. Atmospheric path length and Albedo are also major contributors for high radiance enhancements. The latitude angular dependence is also an important parameter in our calculations to find RE \cite{Siddiqui2017b}.

Table \ref{tab1} shows the RE and CRE values of individual wavelength bands of $\rm{O_2, H_2O, CO_2}$ and $\rm{CH_4}$. The lower values in RE and CRE correspond to thick clouds or forest fire aerosols clouds that occur due to high reflectance. In this study we used selected bunch of space observed values from observation numbers 40 to 52. The observation numbers 43 to 49 are in a good agreement with our RE model for the forest fire detection.

\begin{table}[H]
\centering
\caption{{\sffamily{RE and CRE values for the week 11, pass 69 and observation numbers 40 to 52. $RE_1$, $RE_2$, $RE_3$ and $RE_4$ correspond to wavelength sub-bands of oxygen, water, caron dioxide and methane gases, respectively.}}}\label{tab1}
\scriptsize
\begin{tabular}[t]{lll|lllll}
\hline
{\bfseries{Week}} & {\bfseries{Pass}} & {\bfseries{Obs.\#}} & ${\bf{{RE_1\,(O_2)}}}$ &	${\bf{{RE_2\,(H_2O)}}}$	& ${\bf{{RE_3\,(CO_2)}}}$ &	${\bf{{RE_4\,(CH_4)}}}$ & {\bfseries{CRE}} \\
\hline
11 & 69 & 40 &  0.7214   &  2.344764 &  2.260691 &  0.783366 & 6.110221 \\
11 & 69 & 41 &  0.549205 &  1.764849 &  1.727316 &  0.508873 & 4.550243 \\
11 & 69 & 42 & -0.00688  &  0.753885 &  0.876892 &  0.039664 & 1.663564 \\
11 & 69 & 43 &  0.135247 &  0.429764 &  0.61495  &  0.072215 & 1.252176 \\
11 & 69 & 44 &  0.124571 &  0.213207 &  0.215502 & -0.13654  & 0.416739 \\
11 & 69 & 45 & -0.41421  & -0.12293  & -0.06696  & -0.40725  & -1.01134 \\
11 & 69 & 46 & -0.39236  & -0.11299  & -0.06674  & -0.40589  & -0.97799 \\
11 & 69 & 47 & -0.31398  & -0.04154  &  0.030332 & -0.33553  & -0.66072 \\
11 & 69 & 48 & -0.07514  &  0.165338 &  0.444729 &  0.013554 & 0.548477 \\
11 & 69 & 49 & -0.78618  & -0.30506  & -0.12341  & -0.46567  & -1.68032 \\
11 & 69 & 50 & -0.48266  & -0.11653  & -0.04798  & -0.4106   & -1.05776 \\
11 & 69 & 51 &  0.446414 &  0.882347 &  0.535246 & -0.06945  & 1.794557 \\
11 & 69 & 52 &  0.74834  &  1.229831 &  0.88675  &  0.212886 & 3.077807 \\
\hline
\normalsize
\end{tabular}
\end{table}

In our RE model we incorporate all important parameters as discussed in this section. Figure \ref{fig12} demonstrates the frequency bar chart of aerosol cloud scene due to forest fire (both parameters are unitless). The vertical red line separates forest fire cloud scene reflectance (CRE) with higher reflectance due to other surfaces. Future work requires elaboration and analysis of the CRE values within range on the right side from the red line in order to clarify the nature of the relatively high reflectance.

In the development of this approach we applied the weighted sum based on areal coverage with different canopy constituents. The determination of areal coverage by canopies can be found, for example, by fractional calculus and retinex \cite{Cao2018} that can be used to improve texture information of an image, similar to one shown in Fig. 5.

As a future development we work on RE methods and algorithms that can be used to distinguish water vapor/ice clouds from forest or wildfire clouds without weather forecast and image processing datasets. This can be achieved, for example, by matching not only Albedo but also enhanced column of $\rm{CO_2}$ greenhouse gas concentration due to intense forest or field fires.

\section{Conclusion}

In this work we generalize the Radiance Enhancement method for detection of the cloud scenes to detection of the forest fires. This method can cover a wide spectral range from 1100 nm to 1700 nm by using space observation datasets of Argus 1000 micro‑spectrometer. The RE method can be implemented for detection of combustion aerosols due to high sunlight reflectance of the smoke originated from the forest fires. Our model accounts for the wavelength dependent reflectance and is developed by a new method based on the weighted sum. As the nanosatellite rotates periodically around the Earth, the proposed approach may be a promising technique for continuous monitor of dynamics of the seasonal forest fires.

\section*{Acknowledgments}

This study is supported by Department of Physics and Astronomy at York University, Epic College of Technology, Epic Climate Green (ECG) Inc. and Thoth Technologies Inc. The authors would like to express their gratitude to Dr. Robert Zee and his team from University of Toronto Institute for Aerospace Studies for support, guidance and suggestions in operating of the CanX-2 spacecraft.

\bigskip

\end{document}